\documentclass[11pt,twoside]{article}
\usepackage{asp2010}
\usepackage[colorlinks=true,pdfauthor={A. Accomazzi},pdftitle={Astronomy 3.0 Style}]{hyperref}

\resetcounters

\begin{document}

\title{Astronomy 3.0 Style}
\author{Alberto Accomazzi}
\affil{Harvard-Smithsonian Center for Astrophysics, 60 Garden Street,
  Cambridge, MA 02138, USA}

\begin{abstract}
Over the next decade we will witness the development of 
a new infrastructure in support of data-intensive 
scientific research, which includes Astronomy.
This new networked environment will offer both challenges and
opportunities to our community and has the potential to
transform the way data are described, curated and preserved.
Based on the lessons learned during the development and management of 
the ADS, a case is made for adopting 
the emerging technologies and practices of the
Semantic Web to support the way Astronomy
research will be conducted.
Examples of how small, incremental steps can,
in the aggregate, make a significant difference in the
provision and repurposing of astronomical data are provided.
\end{abstract}

\section{Introduction}

A coming era of networked, data-driven scientific investigation has
been predicted for over a decade now \citep{1998IAUS..179..455S}.  Spurred by 
a stunning increase in the amount of digital data generated
by the current generation of detectors being built or planned
(e.g. Pan-STARRS, LSST, ALMA), scientists
find themselves in the situation of having to gain skills and 
master technologies which have traditionally been the expertise of computer scientists
and engineers.  It has been argued that the coming deluge of data will
generate a paradigm shift in the way scientific research is carried out \citep{4thparadigm}.
Science will require an infrastructure of distributed computational resources
acting upon data scattered on different archives and databases, and
taking advantage of grid or cloud computing resources.
The term
cyberinfrastructure has been proposed to describe this
new digital networked environment, while the term e-Science has been adopted 
to describe the research activity associated with it.

In Astronomy, the need for a new framework supporting this type of data-intensive
research is becoming an urgent matter.  
In order for Astronomy as a discipline to thrive over the next decade,
our research community needs to adopt the cyberinfrastructure framework,
in addition to making
sure that today's graduate students acquire the skills
necessary to take advantage of this environment.
In (partial) response to the first need, the community has created, and
is vigorously funding, Virtual Observatory efforts worldwide.
In response to the latter need,
scientists and archivists are now calling for the creation of the
research fields of Astrostatistics \citep{2010arXiv1004.4148F} and 
Astroinformatics \citep{2009astro2010P...6B}, aiming to bridge
the gap between astronomers and researchers in the field of statistics, data
mining, and semantic computing.

The paradigm shift which is taking place
across most scientific disciplines 
will be heavily influenced by the 
evolution of the Web, which has become the architecture
upon which scientific research is today being conducted.
We are now in the third decade of the development of the Web,
and the term Web 3.0 has been used to
indicate the mix of technologies and practices that are going
to shape its evolution in the coming years.  
While there is no unanimity as to what the Web 3.0 will
look like, most people agree that it will draw heavily on 
the technologies underpinning the so-called Semantic Web
\citep{semweb}, facilitating the creation of applications
enabling intelligent searches, streamlined workflows, 
and highly personalized services.  
Astronomy 3.0 is the term we use in this paper to describe
the research activity that astronomers will carry out in this
environment during the next decade.  It will involve the use
of an ecosystem of interacting web-based resources, including the
infrastructure provided by the
Virtual Observatory, data provisioning services from Astronomy archives,
a variety of analysis services such as Astrometry.net, 
notification services such as skyalert.org, and 
visualization services such as CDS's Aladin and 
Microsoft's WorldWideTelescope.

In this paper we discuss the current status and possible evolution of
the cyberinfrastructure supporting Astronomy 3.0.  Rather 
than attempting to describe an all-encompassing 
view of what the field may look like over the
next decade, we focus on those aspects and technologies that directly
affect the way resources are described, metadata is exchanged,
and applications are built.  We offer examples of how seemingly
small steps in the curation and exposure of metadata can provide
great improvements in the way data are accessed and repurposed.
While much of the scenarios and examples described here draw upon
our experience in the development and maintenance of ADS services, 
we believe that 
the general principles underlying our approach
can be fruitfully applied to a vast number of projects in
Astronomy.

\section{The Astronomy Research Lifecycle}

Because scientific research requires repeatability, it is crucial that all
the (digital) aspects of the scientific lifecycle be properly identified,
annotated, made accessible, and properly linked to each other.  
This includes data, claims on data, processes, and results.  Each of 
these items should be properly modeled and annotated with the relevant
metadata, be it of technical nature (e.g. information about the detector used
to obtain an image) or otherwise (e.g. information about the
program requesting the observation being taken).

\citet{2009arXiv0906.2549P} describe a framework which can be used to represent the 
scientific lifecycle in the era of e-Science.  Recognizing that
all artifacts used in current research are now in digital form
and are often available on the web, they propose a 
web-based model which attempts to formalize the relationships between
these digital assets and capture them as resource aggregations.
According to this model, one has to first identify the components
of the scientific lifecycle, document and curate their metadata,
and then create the proper connections across them.  
The model advocates performing this activity end-to-end, 
starting from the planning
phases, through the data collection and reduction process, and 
ending with the publication of results.  

In the case of Astronomy, the research lifecycle can be 
modeled as consisting of three main phases: science planning,
data acquisition and analysis, and publication of results.  
The science planning phase
typically consists of the formulation of a research goal, the
creation of a science case for it, and possibly the request for
resources such as observing time on a telescope through the submittal 
of one or more proposals.
The activity of writing a proposal or project
plan requesting the allocation of resources is the time when
the project goals, hypothesis and dependencies are described in great
detail, as well as the venue in which technical considerations
about the observing instruments must be taken into account.  It is not
unusual for scientists to run simulations at this stage in order to be able
to predict the expected outcome of observations
based on the observational constraints.
One should note that research per se does not
necessarily require the observation of new data, 
and it is quite possible that a project may 
involve the repurposing and analysis of existing datasets
already available from the Virtual Observatory.
It is, in fact, likely that data re-use will become
the norm rather than the exception in the future.
As a point in case, 
\citet{2009astro2010P..64W} have recently shown that the current use
of archival data from the Hubble Space Telescope exceeds the
use of new data observed by the telescope.

The data acquisition phase involves
the traditional process of obtaining digital data from 
a (virtual or physical) telescope.  This is followed by
a data reduction and analysis step during which data
are processed, normalized, merged with other data sources,
and analyzed for the purpose of exposing their 
characteristics as they relate to the research project goals.
This activity may involve a great deal of
expertise and decision-making on the part of the researcher,
and which can be particularly hard to document and quantify
in an unambiguous way.  
However, it is crucial for scientists to be able to 
reproduce the results of their research, which is dependent on 
the proper documentation of the process followed in this phase.

Finally, the research process usually culminates with the 
publication of the results of this activity.  This typically
involves writing reports, creating
high-level data products synthesizing the characteristics
of the dataset under study, and publishing one or
more scientific papers detailing purpose, methodology,
and findings of the study.  While the electronic publishing
process has made it easier for users to locate
articles of interest, much of the remaining results
created during the research process have traditionally 
not been easily discoverable.
High-level data products published as tables and 
images within the articles are often difficult to locate
as digital objects because they are not usually
tagged and indexed with the electronic paper in an efficient way.

In the case of Astronomy, there are now a number of well-established
projects focusing on the curation of different aspects of
this lifecycle.  The main players today are 
the publishers and ADS (for 
bibliographies), NED and SIMBAD (for object metadata),
Vizier (for electronic catalogs), 
and a number of distributed archives (for primary data
products such as images and spectra).
Virtual Observatory projects provide the tools and protocols to
easily access the data and metadata available from all these
different archives in a consistent way.  However, it is still
the case that only a fraction of the data and metadata 
generated by a research project is captured and made
available on the web today.  Some of this content ends up
being interlinked with the rest of the resources related to it
(e.g. the way the record for a paper is linked to the objects
mentioned in it), but since the creation of links is an activity requiring
expert curation, it depends entirely on the efforts of 
several distributed projects.  Even less of
this on-line content today is linked in a way that provides us with an efficient
way for applications to take advantage of it.
In the next section we will review the status of this
infrastructure and offer a vision of how it should be 
improved in order to take advantage of the Web 3.0 environment.

\section{The Web of Astronomy Links}

Astronomy was one of the first disciplines to take advantage of
the early developments of the web
\citep{1994AAS...185.4102A}.  As early as 1993 it became
possible to perform a literature search in ADS in conjunction with 
an object query in SIMBAD, as the result of a collaboration between
the two projects.  Article records in ADS were linked to object 
records in SIMBAD, and vice-versa, allowing users to seamlessly
move across the two databases and explore the relationships
between their data holdings.
In 1997 links were created between ADS records and observational
data available from the main NASA archives as well as ESO.
Thus, authors could easily access observations which had been
studied in a paper, and conversely, access all publications
which referenced a particular dataset.

The fact that agreements allowing the creation of these links 
were established so quickly highlights the desire of the 
community to fully exploit the technological advantages offered
by the web in the early 1990s.  An agreement between the
data centers codifying a system to uniquely identify bibliographic
records \citep{1995VA.....39R.272S} allowed all interested parties
to unambiguously compute such identifiers based on their metadata,
as well as create links between them and other resources.  
The introduction of such identifiers (bibcodes) took place ten years 
before publishers agreed on a system to uniquely identify
articles via the DOI (Digital Object Identifier) system, which 
shows that interoperability within a particular community
can be achieved quickly and successfully when necessary.
Of course, interoperability between bibcode identifiers
and the DOI system is now maintained by projects such as ADS in
a transparent way, so that one does not need to choose one
system over the other.

The adoption and mutual exchange of these links has benefited
the astronomical community at large, starting from end-users
who can now click their way through this network of research
data, and including the archives themselves, which have
seen a significant rise of use due to their greater 
interconnectedness
(as an example, in 2009, the ADS recorded over 100,000
clicks from its bibliographic records to data products hosted
by external archives).
Beginning in the late 1990s, libraries 
began playing an important role in maintaining links
between bibliographies and data products.  Several institutions
today use ADS as a search tool to keep lists of bibliographies
related to their missions and share some of this metadata back
with ADS.  This allows the possibility of searching the literature
with a filter limiting results to the contributions of a
particular institution.  Thanks to this synergy, metadata that
librarians have started collecting for the main purpose of 
generating reports and maintaining metrics can now be used
by ADS to enhance literature searches in different ways.
For instance, due to the contributions of the librarian
from the Space Telescope Science Institute and the archivist from 
the Chandra X-ray observatory,
one can now use ADS to find papers on a particular topic
(e.g. ``globular clusters'')
that have optical data from the HST and X-ray data from 
Chandra.  This example shows how the 
creation, use, and repurposing of links between
metadata records held by different projects on different web sites can
enable new discoveries when the proper connections between them are made
explicit.

The amount of current links, or connections, between pieces of metadata
describing online astronomical resources today is quite
impressive.  The figures in Table 1 describe 
a subset of the number
of links existing between ADS bibliographic records and
other ADS records (internal links), or resources available
from other archives (external links) as of 
April 2010.  Considering that the total current number
of records in the ADS databases is approximately 8 million, one can
see that the number of connections between them and 
other resources is one order of magnitude larger.
Harnessing these connections can provide new ways to view
and interpret the network of linked resources.  For instance, ADS 
uses its citation and co-readership network to generate recommendations
to individual readers, both on an article-by-article basis and
in the aggregate, through its myADS notification service
\citep{2003AAS...203.2005K}.

\begin{table}[!ht] 
\caption{ADS Link Statistics: a selected list of the links
between bibliographic records and other resources.  Units are
millions (M) or thousands (K).  Links are categorized as either
being internal (i.e. pointing to other ADS records) or external
(i.e. pointing to resources on other websites).}
\smallskip 
\begin{center} 
{\small 
\begin{tabular}{lrl}

\tableline 
\noalign{\smallskip} 
Link & Count & Type\\
\noalign{\smallskip} 

\tableline 
\noalign{\smallskip} 
citations & 40M & internal\\
co-readership & 18M & internal\\
fulltext & 5M & internal \& external \\
astronomical objects & 250K& external \\
data products & 130K & external\\
bibliographic groups & 200K & internal \\
\noalign{\smallskip} 
\tableline 
\end{tabular} 
} 
\end{center} 
\end{table} 

\citet{2009arXiv0912.5235K} and \citet{2010APS..MAR.S1244H} have been exploring the
use of these networks to build a recommender system based on
citation and co-readership data.  Such a system would be similar in
scope to what some commercial websites now offer to their clients.
For example, by analyzing purchase histories and user access to product
information, Amazon is able to suggest products that might be 
of interest to a shopper.
In the world of scholarly literature, connections based on
co-authorship, citations, keyword co-occurrence, readership,
links to similar data products, etc. provide the basis for
recommendations.
In the remainder of this paper we will present a framework
that can be used to model the metadata underlying such a system and
we suggest ways in which we, as a community, can 
begin to adopt some of its practices in an incremental way.

\section{The Semantic Web}

While it is relatively simple for projects to exploit the connections
between resources under their own control in order to create new applications, 
making use of the links between resources maintained on different sites
is a more complex matter.   For the most part, the task of interpreting the
connections found in the distributed network of Astronomy resources
has so far been left to the end-users.  Therefore, the knowledge that is embedded
in this network and that could be gained through the analysis
of its topography has so far remained untapped.

Tim Berners-Lee, the inventor of the World Wide Web, has long been advocating
for the creation of the Semantic Web \citep{semweb}, a networked
environment in which meaning is attached to resources available on the web
so that both humans and machines can make use of it.  Two
fundamental components of the Semantic Web are its use of
ontologies to represent concepts and the relationships between them
and its reliance on a linked data model.
The Linked Data effort \footnote{\url{http://www.linkeddata.org}}
is based on a few simple principles:
(1) resources are named via HTTP URIs;
(2) metadata is open and in a standard format (RDF);
(3) resources are interlinked and their links are typed.
Taken as a whole, these guidelines describe how exposing metadata and
links between them can be used today to build a global graph of 
resources built on the architecture of the web.  This means,
among other things, that it becomes simple for both
people and applications to 
use the metadata and links in order to transverse, analyze, and compute
over this graph.

The use of a linked-data approach to describe and connect 
resources imposes some requirements on information providers
but offers several major advantages to the community.  
Primarily, it requires discipline on the part of
data archives and forces them to identify the important
pieces of data holdings that they serve.  Once these resources have
been identified, they need to be uniquely named (via URIs), and
their metadata must be exposed in a machine readable format 
(RDF).  Mappings between identifiers can be formally expressed
when necessary (for instance, to indicate that a particular DOI 
and bibcode correspond to the same article), and allow 
applications to unambiguously gather information about resources.
Finally, relationships between different resources should be included 
in the form of typed links (for example, linking an article to the set
of observations available in an online archive).
We believe that the advantages this approach can provide to the community
greatly outweigh the amount of effort involved.  In the next section
we explore some of the applications that will be enabled
by this model.

\section{Semantic Astronomy Applications}

In the long term, emerging technologies and practices adopted
in the creation and evolution of the Semantic Web will likely
underpin the scientific cyberinfrastructure of Astronomy 3.0.
However, even before such a unified ``web of data'' 
linked together by relationships described by well-defined ontologies 
emerges, much work can be accomplished by incrementally exposing
well-structured metadata from existing repositories.  
As an example of what archives and libraries can do today to enhance
the repurpusing and increase the use of their holdings by adopting 
best practices advocated by the semantic web proponents,
consider providing enhanced records containing machine-readable
metadata.
The ADS has been supporting this effort by adopting best practices emerging
from the Digital Library world as well as from the field of Astronomy.  For instance, 
since 2009, all the ADS HTML abstract pages now contain machine-readable
metadata describing the record in question (see figure 1).  This
is in addition to the existing metadata output formats that ADS has been
providing since 2002 (various flavors of XML, including Dublin Core,
EndNote, and RIS).

\begin{figure} [!ht]
\includegraphics[width=1.0\textwidth]{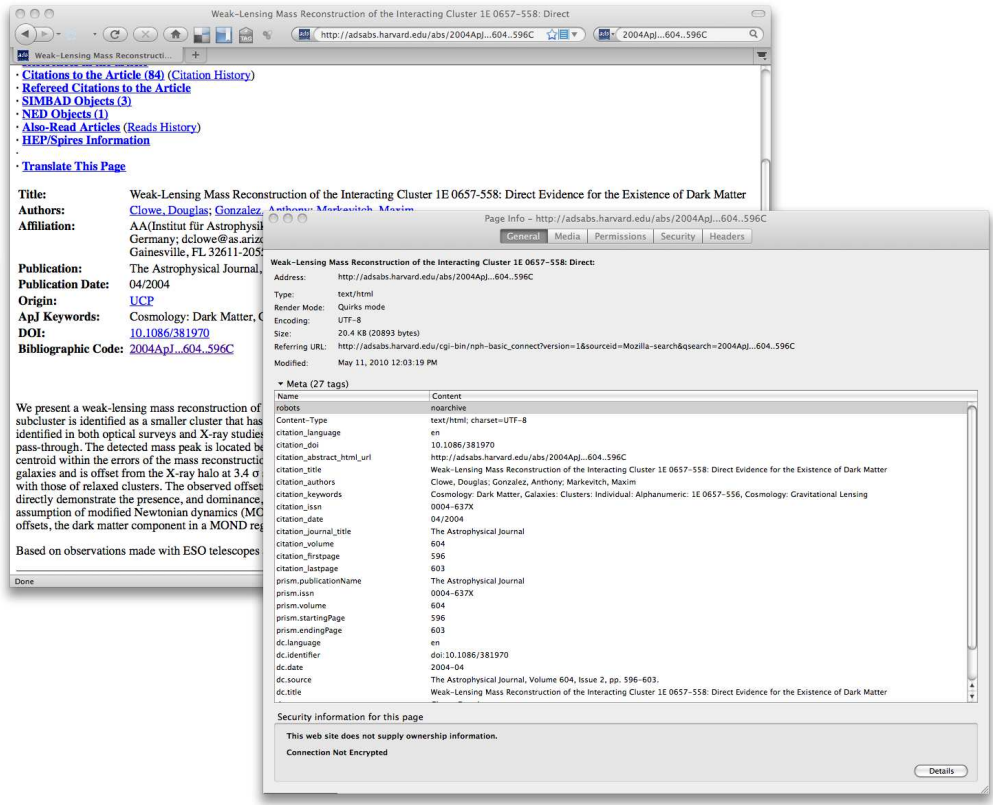} 
\caption{A listing of the metadata embedded in a traditional HTML
page for an ADS bibliographic record, as displayed by Firefox 3.}
\label{fig1} 
\end{figure}

While the inclusion of these tags in the HTML header does not change
the look of the page to the end user, it provides a mechanism for
crawlers and software agents to
programmatically extract and use resource metadata
from a normal-looking HTML page.  Having a single
URI uniquely identify a resource and providing actionable metadata
to both users and machines is not only a convenient way to expose
information about the contents of a database, but it is also considered a good
design choice since it is based on the very architecture of the web.  
One of the immediate advantages of following such a 
practice is that it greatly facilitates the work of crawlers.
A tangible result of the embedding of metadata in our HTML pages
has been the efficient indexing of ADS records in the major search
engines and Google Scholar, which now account for over half of all 
accesses to our databases.

Similarly, small, incremental steps taken by individual data providers can, in 
the aggregate, make a significant difference in the provision and repurposing
of data and metadata.  
Astronomy librarians, in their role of maintainers of 
bibliographic collections
and metrics related to their institutes' data products, 
have a significant role to play in this effort.  
By sharing more of the metadata that they currently collect as part of 
their institutional bibliography collection, they can provide
ADS and similar projects valuable observational metadata
required to create applications that make use of metadata 
integrated from different archives (optical, X-ray, radio, etc.)
and that describe different resources (bibliographies, 
observations, objects, etc.).

As an example, consider the prototype interface recently developed
by the ADS which makes use of a literature search combined with
a search on astronomical object metadata.
A researcher may wish to know which astronomical objects are
most often referenced in review papers about ``weak gravitational lensing.''  
This question can be answered by creating views, or ``facets,''
of the literature in question based on astronomical objects.
Figure 2 shows the interface that ADS has recently
implemented by combining the traditional ADS topic search with
a ranked list of objects appearing in the list of results 
as returned by the SIMBAD database.  This ranked list provides
a set of facets that can 
be used to filter or further select and rank the search results.
Using this additional information, the researcher can quickly 
identify which objects are most relevant to the topic of interest
and access information about the objects or retrieve the papers
that mention them.

\begin{figure}[!ht]
\includegraphics[width=1.0\textwidth]{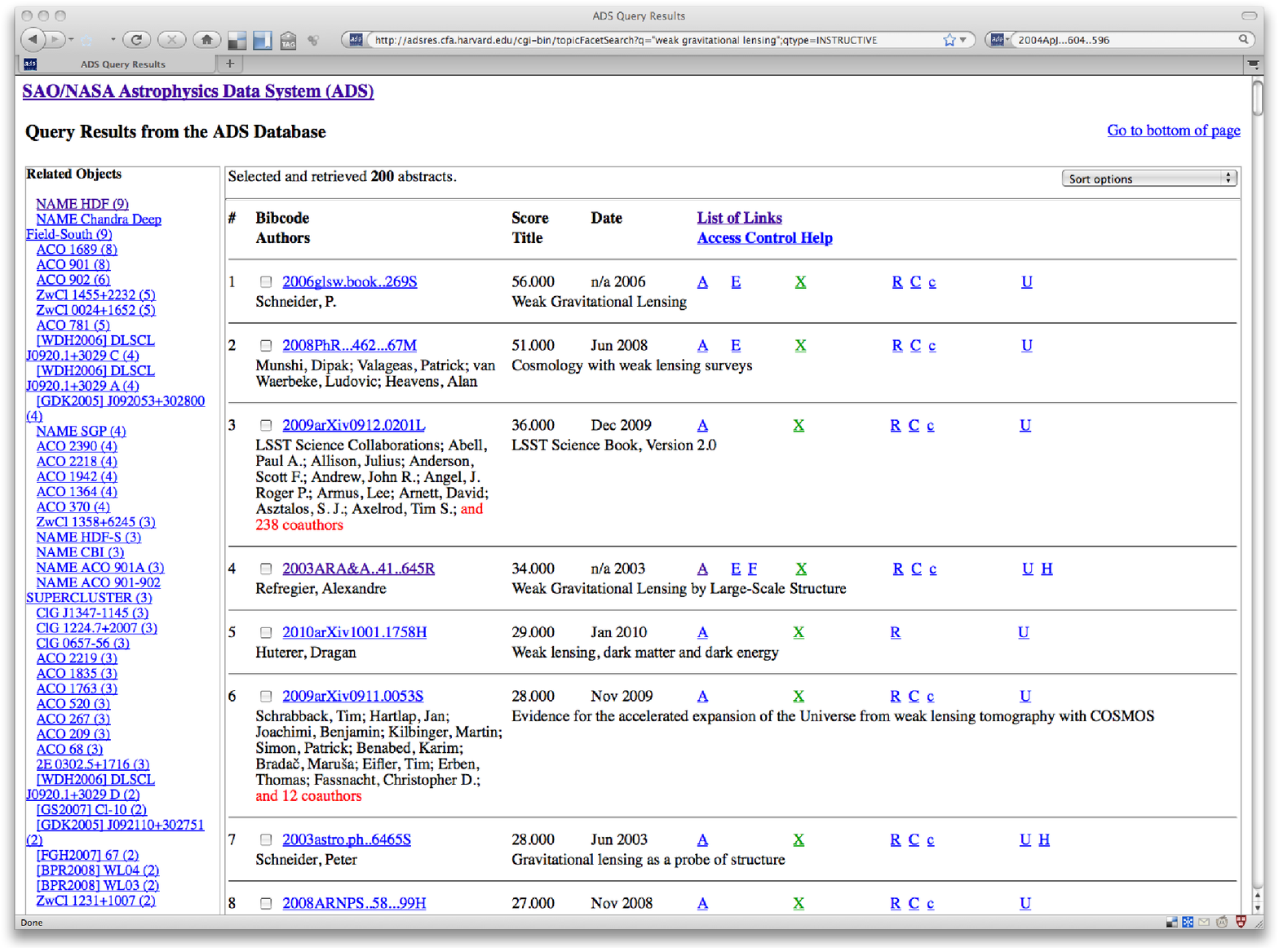} 
\caption{A prototype application implementing object-based facets as a
view into search results.  The original search requested review
articles about ``weak gravitational lensing.''  
The list of names displayed on the left bar represent the most
referenced objects mentioned in the literature on the subject.}
\label{fig2} 
\end{figure}

It is easy to imagine a variety of similar scenarios in which 
bibliographic or observational 
properties are used to create similar facets. For example, when manipulating a list of papers, facets 
could be built on the sets of keywords that are shared amongst them, or the type of data products 
associated with them. While not all facets may make sense in all cases, there are certainly several 
scenarios in which they provide insightful views into the data. 
We plan to investigate the usefulness and impact of such
interfaces to enable more advanced views of the metadata describing
astronomical resources.  For instance, one may want to ask ``what
are the most cited papers discussing objects in this spectral band
and in this area of the sky.''  Answering this type of question 
requires connecting pieces of metadata which currently exists but
which are not linked in an efficient, machine-readable way.

\section{Conclusions}

Particularly in this era of data-intensive research,
integrating and manipulating pieces of information from
different sources can provide a new context for the 
analysis and evaluation of the phenomena behind them.
Important aspects of this information may become
apparent after crucial connections are made
and new views based on them are created,
exposing evidence which might have otherwise 
been missed in the sheer volume of data.
In this paper we have outlined a model for describing the lifecycle of 
astronomy research in the era of the Web 3.0.  This model
advocates for the preservation of all artifacts and workflows
generated during the research activity.
We have argued for the adoption of emerging technologies 
in use in the Semantic Web to formally describe these
resources, their aggregations and relationships.
This model is particularly appropriate for our field due to the
distributed nature of the curation activities which take
place during the research lifecycle in Astronomy.

However, technology alone will not provide
solutions to problems requiring community buy-in and
changes in policies at the institutional level.
\citet{2010arXiv1005.1886A} have recently advocated 
for community support of such an effort.  In order for it to 
be successful, broad participation is required from all the
stakeholders involved in the curation and preservation of research
data in Astronomy.  This includes data archivists, librarians,
publishers, and scientists.  Observatories, institutes, and 
projects which have adopted an open-sky policy for their data
holdings should now consider embracing an open-metadata
policy to allow Astronomy 3.0 to develop and deliver its full
potential to the community.

\acknowledgements This work was supported by the Astrophysics Data
System project which is funded by NASA grant NNX09AB39G.  The author
thanks Michael Kurtz for his comments and suggestions and the
Editors for their patience.

\end{document}